\documentclass[10pt,superscriptaddress,aps,prx,twocolumn,nofootinbib]{revtex4-2}
\usepackage{amsmath,amssymb,amsthm,amsfonts}
\usepackage[breaklinks=true,colorlinks,citecolor=blue,linkcolor=blue,urlcolor=blue]{hyperref}
\usepackage[pdftex]{graphicx}
\usepackage{braket}
\usepackage{comment}
\usepackage{multirow}
\usepackage{siunitx}
\usepackage{float}
\usepackage{algorithm}
\usepackage{algpseudocode}
\usepackage{mathtools}
\usepackage[normalem]{ulem}
\usepackage{tabularx, booktabs}
\usepackage{color,soul}
\usepackage{url}
\usepackage{orcidlink}
\usepackage{soul}

\newcommand{\eqlabel}[1]{Eq.~\eqref{#1}}
\newcommand{\figlabel}[1]{Fig.~\ref{#1}}

\newcommand{\tablabel}[1]{Table~\ref{#1}}

\begin{document}

\title{Hybrid Sequential Quantum Computing}
\author{Pranav Chandarana$^{\orcidlink{0000-0002-3890-1862}}$}

\affiliation{Kipu Quantum GmbH, Greifswalderstrasse 212, 10405 Berlin, Germany}
\affiliation{Department of Physical Chemistry, University of the Basque Country EHU, Apartado 644, 48080 Bilbao, Spain}

\author{Sebastián V. Romero$^{\orcidlink{0000-0002-4675-4452}}$}
\affiliation{Kipu Quantum GmbH, Greifswalderstrasse 212, 10405 Berlin, Germany}
\affiliation{Department of Physical Chemistry, University of the Basque Country EHU, Apartado 644, 48080 Bilbao, Spain}

\author{Alejandro Gomez Cadavid$^{\orcidlink{0000-0003-3271-4684}}$}
\affiliation{Kipu Quantum GmbH, Greifswalderstrasse 212, 10405 Berlin, Germany}
\affiliation{Department of Physical Chemistry, University of the Basque Country EHU, Apartado 644, 48080 Bilbao, Spain}

\author{Anton Simen$^{\orcidlink{0000-0001-8863-4806}}$}
\affiliation{Kipu Quantum GmbH, Greifswalderstrasse 212, 10405 Berlin, Germany}
\affiliation{Department of Physical Chemistry, University of the Basque Country EHU, Apartado 644, 48080 Bilbao, Spain}

\author{Enrique Solano$^{\orcidlink{0000-0002-8602-1181}}$}
\email{enr.solano@gmail.com}
\affiliation{Kipu Quantum GmbH, Greifswalderstrasse 212, 10405 Berlin, Germany}

\author{Narendra N. Hegade$^{\orcidlink{0000-0002-9673-2833}}$}
\email{narendrahegade5@gmail.com}
\affiliation{Kipu Quantum GmbH, Greifswalderstrasse 212, 10405 Berlin, Germany}

\date{\today}
\begin{abstract}
We introduce hybrid sequential quantum computing (HSQC), a paradigm for combinatorial optimization that systematically integrates classical and quantum methods within a structured, stage-wise workflow. HSQC may involve an arbitrary sequence of classical and quantum processes, as long as the global result outperforms the standalone components. Our testbed begins with classical optimizers to explore the solution landscape, followed by quantum optimization to refine candidate solutions, and concludes with classical solvers to recover nearby or exact-optimal states. We demonstrate two instantiations: (i) a pipeline combining simulated annealing (SA), bias-field digitized counterdiabatic quantum optimization (BF-DCQO), and memetic tabu search (MTS); and (ii) a variant combining SA, BF-DCQO, and a second round of SA. This workflow design is motivated by the complementary strengths of each component. Classical heuristics efficiently find low-energy configurations, but often get trapped in local minima. BF-DCQO exploits quantum resources to tunnel through these barriers and improve solution quality. Due to decoherence and approximations, BF-DCQO may not always yield optimal results. Thus, the best quantum-enhanced state is used to continue with a final classical refinement stage. Applied to challenging higher-order unconstrained binary optimization (HUBO) problems on a 156-qubit heavy-hexagonal superconducting quantum processor, we show that HSQC consistently recovers ground-state solutions in just a few seconds. Compared to standalone classical solvers, HSQC achieves a speedup of up to $700\times$ over SA and up to $9\times$ over MTS in estimated runtimes. These results demonstrate that HSQC provides a flexible and scalable framework capable of delivering up to two orders of magnitude improvement at runtime quantum-advantage level on advanced commercial quantum processors.
\end{abstract}

\maketitle

\section{Introduction}

Machine learning, classical optimization, and classical simulation methods build a coupled toolkit for modern science, each with distinct strengths and limitations that unveil what is computationally feasible to date. Machine learning converts input data into predictive or generative models, relevant to biology~\cite{abramson2024accurate}, material science~\cite{merchant2023scaling}, weather forecasting~\cite{lam2023learning}, and conversational agents~\cite{openai2024gpt4}, among others. However, it often requires solving vast high-dimensional problems or it faces trainability issues due to barren plateaus, overfitting, or complex landscapes~\cite{beyer1999when,nakkiran2020deep,zhang2021understanding}. Given a problem encoded as an objective function subject to constraints, classical optimization aims to find the set of optimal variables that minimize or maximize the objective function. With applications covering a broad set of use-cases like logistics~\cite{dijkstra1959note,hart1968formal,coffman2013bin,duan2025breaking} and manufacturing~\cite{liu2018chain,xiong2022survey,sundermann2023evaluating}, these problems often belong to the NP-hard class, demanding an exponentially growing computational time with the problem size, heuristics, and decomposition strategies. Classical simulations of physical systems range from Monte Carlo methods~\cite{metropolis1949montecarlo,metropolis1953equation,barbu2020monte}, which efficiently sample equilibrium ensembles but with slow convergence, as well as density functional theory~\cite{hohenberg1964inhomogeneous,kohn1965selfconsistent,perdew1996generalized,jones2015density}, which investigates the electronic structure of many-body systems by first-principles but struggles with strongly correlated systems~\cite{georges1996dynamical,cohen2008insights}.

On the other hand, recent developments in quantum technologies are turning quantum computers into reliable devices capable of addressing computationally complex tasks, which may overcome the issues aforementioned and ultimately solve problems beyond the reach of classical computation. Apart from purely quantum algorithms, several implementations have been put forth combining a classical precomputation followed by a quantum routine, aiming for a beneficial interplay between both computing paradigms. Some examples are given by a classical warm-start of quantum optimization routines~\cite{egger2021warmstarting}, classical preparation of initial states for ground state preparation of chemical compounds~\cite{peruzzo2014variational,berry2025rapid}, and hybrid decomposition pipelines~\cite{dwave2017partitioning,zhou2023qaoa,naghmouchi2024mixed,acharya2025decomposition}, among others. There are also recent proposals reversing the order, where a short quantum precomputation is followed by a more intense classical routine, sometimes referred to as \emph{quantum accelerators}~\cite{dwave2020hybrid,montanaro2020quantum,peng2025hybrid,jeong2025quantumenhanced}. However, to our knowledge, none of these classical-to-quantum or quantum-to-classical approaches has been able to show any source of runtime quantum advantage to date.
\begin{figure*}[t]
    \centering
    \includegraphics[width=1.0\linewidth]{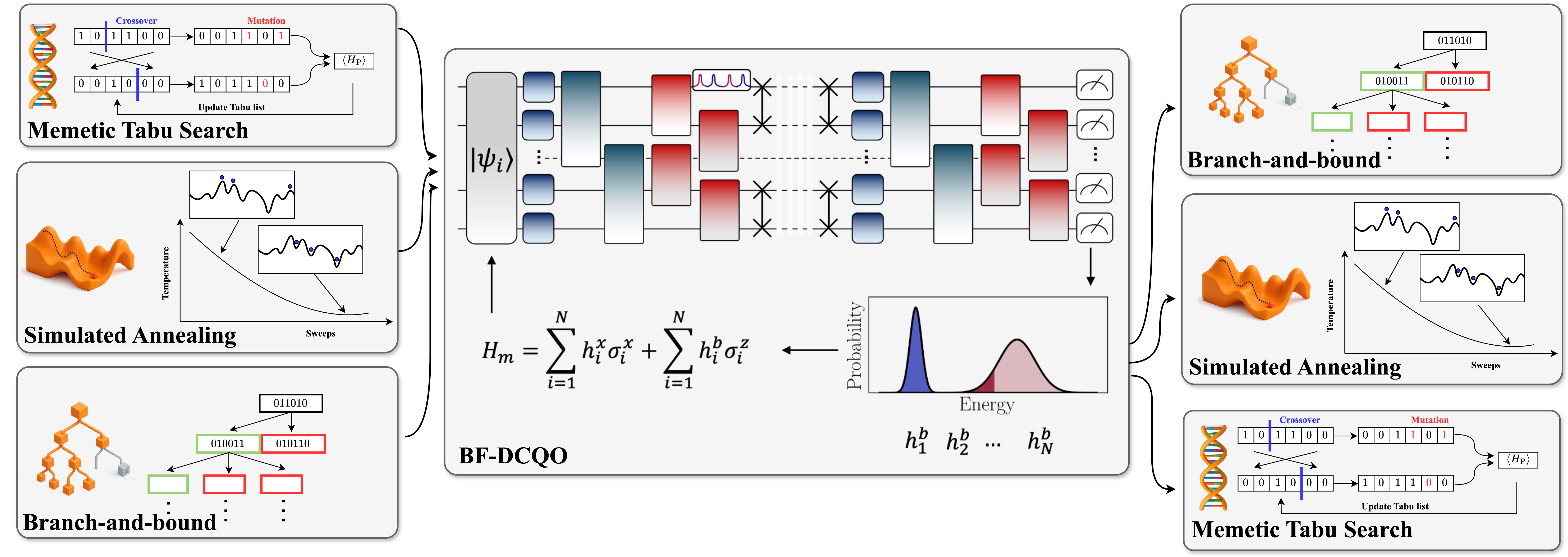}
\caption{\textbf{Schematic of the hybrid sequential quantum computing (HSQC) framework.}
The optimization begins by formulating the target problem as a higher-order unconstrained binary optimization (HUBO) problem. A classical optimization routine first explores the energy landscape to identify promising low-energy configurations, continuing until further improvements become computationally expensive. The best configuration is then used to initialize the BF-DCQO protocol. The QPU then executes a controlled counterdiabatic evolution that guides the system through the energy landscape by exploiting quantum superposition and tunneling to escape local minima and converge toward lower-energy states. The resulting output is passed to a final classical optimization stage, which may employ heuristic or exact methods, to reach the ground state or obtain high-quality approximations. The runtime allocation across the stages is chosen adaptively based on the problem characteristics.
}
    \label{fig:schematic}
\end{figure*}

In this work, we introduce hybrid sequential quantum computing (HSQC), a paradigm that leverages a selective combination of quantum and classical computing routines to solve combinatorial optimization problems where standalone methods struggle (see~\figlabel{fig:schematic}). Classical algorithms are well established, yet many require resources that grow rapidly, often exponentially, to obtain accurate or practically valuable solutions, with some problems remaining out of reach. Quantum computing offers a potential remedy, but current devices are limited by gate and readout fidelities, coherence times, and hardware connectivity. HSQC targets this gap by selectively invoking each routine where it is strongest, combining their advantages and mitigating limitations, in line with recent proposals that combine different platforms~\cite{sqc}. Motivated by recent quantum speedup claims in optimization~\cite{durr1999quantum,somma2008quantum,wocjan2008speedup,hastings2018shortpathquantum,montanaro2018quantumwalk,montanaro2020quantum,chakrabarti2022universal,pirnay2024superpolynomial,boulebnane2024solving,munozbauza2025scaling,chandarana2025runtimequantumadvantagedigital} and without loss of generality, we experimentally validate HSQC by solving higher-order unconstrained binary optimization (HUBO) problems by sandwiching the bias-field digitized counterdiabatic quantum optimization (BF-DCQO) algorithm~\cite{cadavid2024bias,romero2024bias, chandarana2025runtimequantumadvantagedigital, iskay} on IBM quantum hardware~\cite{ibm} between different classical solvers, namely simulated annealing (SA)~\cite{kirkpatrick1983optimization} and memetic tabu search (MTS)~\cite{silva2021quadratic}. For completeness, we also solve them using D-Wave quantum annealers~\cite{dwave} for different annealing times as well as their hybrid solvers~\cite{dwave2020hybrid}, both requiring extra qubits to quadratize higher-order terms. Our results show a significant speedup in both scenarios, against the involved solvers running independently, and the D-Wave solvers. From the results drawn, HSQC is capable of boosting well-established classical routines by providing faster and better solutions using current quantum hardware, with potential applications in quantum machine learning, material simulation, and quantum chemistry, apart from quantum optimization.

\section{Results}\label{sec:Results}
\subsection{Higher-order binary optimization} 
HUBO extends quadratic unconstrained binary optimization (QUBO) by allowing interactions among up to \(p\) binary variables. Many academic and industrial optimization problems can be formulated as HUBO problems~\cite{PhysRevApplied.12.014004}. The cost function is
\begin{equation}\label{eq:hubo_cost}
F(z) \;=\;
\sum_{r=1}^{p}
\;
\sum_{(a_1,\dots,a_r)\in \mathbb{P}_k}
W_{a_1 \dots a_r}\,
z_{a_1}\cdots z_{a_r},
\end{equation}
where \(z_a\in\{0,1\}\), $\mathbb{P}_k$ is a hypergraph containing the indices of the $k$-body terms, \(p\) is the maximum interaction order, and the couplings \(W_{a_1\dots a_r}\) are drawn from a specified distribution.

Instance difficulty is governed by four factors: the variable count \(n\); the statistics of the couplings \(W_{a_1\dots a_r}\); the highest interaction order \(p\); and the interaction density of the underlying hypergraph.  We consider up to \(n = 156\) variables on IBM superconducting processors~\cite{ibm}, fixing \(p = 3\) for hardware compatibility.  Mapping each bit to a spin via \(z_a = \tfrac{1-\sigma_a^{z}}{2}\) converts Eq.~\eqref{eq:hubo_cost} into the problem Hamiltonian
\begin{equation}\label{eq:hubo_ham}
\begin{split}
H_{\mathrm{P}} &=
\sum_{a=1}^{N} h_a\,\sigma_a^{z}
+
\sum_{(u,v)\in \mathbb{P}_{2b}} J_{uv}\,\sigma_{u}^{z}\sigma_{v}^{z} \\
&+
\sum_{(u,v,w)\in \mathbb{P}_{3b}} K_{uvw}\,\sigma_{u}^{z}\sigma_{v}^{z}\sigma_{w}^{z},
\end{split}
\end{equation}
whose ground state minimizes \(F(z)\), with $\mathbb{P}_{\{2b,3b\}}$ containing the indices of the two- and three-body terms, and $N$ the number of qubits. Since our mapping from variables to qubits is one-to-one, we have that $n=N$.

To create challenging yet hardware‐compatible instances, we follow the graph‐coloring strategy of Refs.~\cite{kim2023evidence,sqc}: nearest‐neighbor two- and three-body couplings are first assigned on the native topology, then a \textsc{swap} layer is applied; this process is repeated for \(n_{\mathrm{swap}}\) layers (see Methods). The resulting dense interaction maps for \(H_{\mathrm{P}}\) yield HUBO problems that are difficult for classical solvers while embedding efficiently on today’s quantum devices.

To evaluate the effectiveness of the HSQC strategy, we consider a suite of challenging HUBO instances defined on $N = 156$ qubits. Each instance incorporates $n_{\rm{swap}} = 2$ layers (see Methods), resulting in a problem Hamiltonian $H_{\rm{P}}$ that includes 156 one-body terms, 125 two-body terms, and 616 three-body terms. The coupling coefficients $h$, $J$, and $K$ are sampled from a Cauchy distribution, which has been shown to produce classically hard instances, particularly for algorithms like  SA~\cite{chandarana2025runtimequantumadvantagedigital}. 

As a concluding remark, D-Wave quantum annealers are well-suited for solving QUBO problems formulated as two-body spin glasses (\eqlabel{eq:hubo_cost} setting $p=2$), since their analog hardware naturally evolves according to a transverse-field Ising model. Therefore, in order to solve HUBO instances using their devices, a HUBO-to-QUBO mapping is required, demanding extra qubits and introducing constraints (weighted by a tunable Lagrange multiplier) that are not naturally present in the original HUBO formulation. Additionally, once the Hamiltonian is transformed, a suitable embedding across the coupling map of the device has to be found, which might increase the final number of qubits required. Note that in this case, $n\neq N$. In our experiments, the first conversion is done using \textsc{dimod} library and the subsequent embedding on hardware is done using the \textsc{D-Wave Ocean SDK}~\cite{ocean} which, given a minor and a target graph, heuristically attempts to map the minor into the target, where additional qubits may be required to finally embed the resulting QUBO problem. A more detailed analysis can be seen in Methods.

\subsection{SA + BF-DCQO + MTS}
\label{sec:mts}
\begin{table}[t]
    \caption{\textbf{Classical hardware and software specifications.}}\label{tab:specs}
    \begin{ruledtabular}\begin{tabular}{rlrl}
       Processor & \multicolumn{3}{l}{AMD (KVM processor)  ($48\times\SI{2.3}{GHz}$)} \\
       RAM & \multicolumn{3}{l}{$\SI{123}{GB}$} \\
       OS & \multicolumn{3}{l}{Debian GNU/Linux 12 (bookworm) ($\times 64$)} \\
       CPLEX~\cite{cplex} & v22.1.2.0 & C++ & 11.4.0 \\
    \end{tabular}\end{ruledtabular}
\end{table}
\begin{figure}[t]
    \centering
    \includegraphics[width=\linewidth]{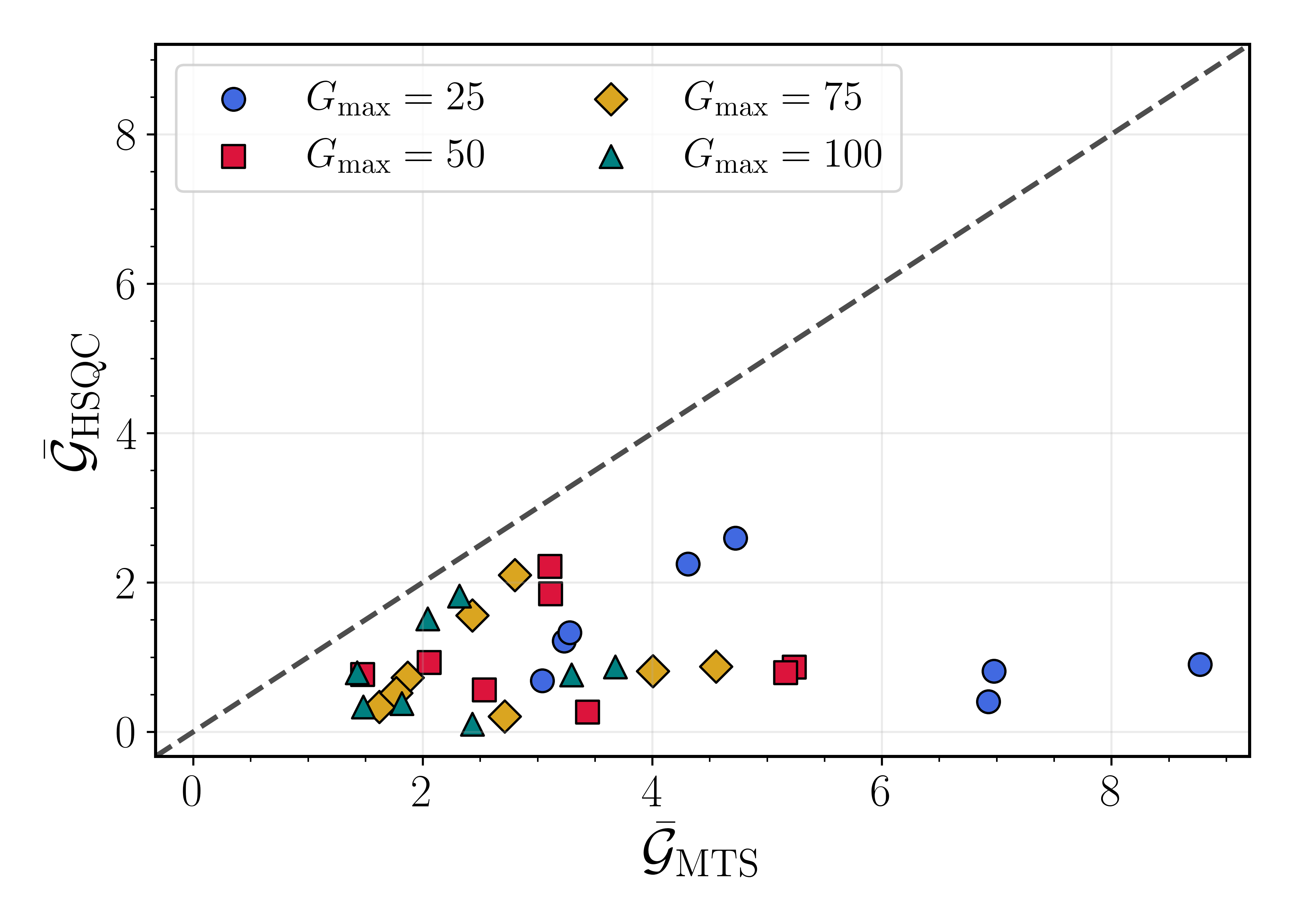}
    \caption{\textbf{Performance comparison between MTS and HSQC.} For different number of generations $G_{\max}$, optimality gaps $\mathcal{G}$ obtained across the eight instances tested in~\tablabel{tab:mts_bfdcqo}. Data points lying in the lower part of the plot mean that HSQC approach performed better than MTS.}
    \label{fig:AR-MTS}
\end{figure}
In this section, we present a 
HSQC strategy that combines three optimization solvers:
SA,
BF-DCQO, and 
MTS.
SA~\cite{kirkpatrick1983optimization} is a stochastic optimization method inspired by the process of physical annealing. It explores the energy landscape by accepting both downhill (energy-lowering) and, with decreasing probability, uphill (energy-increasing) moves. This feature enables the algorithm to escape local minima and converge toward near-optimal solutions. Within the HSQC framework, SA is executed first over $n_{\text{runs}}$ independent trials, each consisting of $n_{\text{sweep}}$ sweeps (see Methods).

Continued from the best SA solution, the second stage of HSQC applies BF-DCQO, a digitized quantum algorithm based on counterdiabatic (CD) protocols \cite{hegade2021shortcuts}. The algorithm implements a fast, discrete-time quantum evolution governed by a CD Hamiltonian $H_{\text{cd}}$, which interpolates between a mixer Hamiltonian $H_m$ and a problem Hamiltonian $H_{\text{P}}$. In the bias-field variant~\cite{cadavid2024bias,romero2024bias,iskay}, the mixer Hamiltonian is dynamically updated based on measurement-driven feedback
%\begin{equation}
$H_m = \sum_{i=1}^N \left(h^x_i \sigma^x_i + h^b_i \sigma^z_i\right)$, 
%\end{equation}
where the transverse fields $h^x_i$ are fixed, while the longitudinal bias fields $h^b_i$ are iteratively adjusted using the rule $h^b_i = \pm\langle \sigma^z_i \rangle$. This iterative feedback loop is executed for $n_{\text{iter}}$ steps, progressively improving the initialization of the quantum state prior to evolution (see Methods).

The final component of the HSQC protocol is MTS, a memetic metaheuristic that merges evolutionary algorithms~\cite{https://doi.org/10.1002/widm.1124} with local search guided by a tabu list~\cite{tabu}. Starting from an initial population of $P$ candidate bitstrings, constructed by cloning the best solution $\mathbf{s}_{\text{ws}}$ obtained from BF-DCQO, the algorithm proceeds through $G_{\max}$ generations. In each generation, pairs of parents are selected, recombined via crossover, and mutated with a generation-dependent mutation rate
\begin{equation}\label{eq:decaying_mutation}
\mu_g = \mu_{\mathrm{end}} + (\mu_{\mathrm{start}} - \mu_{\mathrm{end}})\frac{\ln(G_{\max} + 1 - g)}{\ln(G_{\max} + 1)},
\end{equation}
where $g$ denotes the current generation. Each offspring is further refined via a local tabu search before being reintroduced into the population.

The complete HSQC workflow proceeds as follows. First, SA is run for $n_{\text{sweep}}$ sweeps across $n_{\text{runs}}$ trials, resulting in a classical runtime of $T_{\text{SA}} = n_{\text{sweep}} \times n_{\text{runs}} \times 0.6 \times 10^{-5}$ seconds. The lowest-energy configuration from SA is used to initialize the bias fields $h^b_i$ for the subsequent BF-DCQO stage. BF-DCQO is executed on quantum hardware for $n_{\text{iter}}$ iterations and $n_{\text{shots}}$ shots, with total runtime given by $T_{\text{BFDCQO}} = T_{\text{CPU}} + T_{\text{QPU}}$. The quantum sampling time is set to $T_{\text{QPU}} = n_{\text{shots}} \times 10^{-4}$ seconds~\cite{kotil2025quantum}, and each iteration includes a classical local search phase lasting $T_{\text{CPU}} = n_{\text{sweep}}^{\text{loc}} \times 0.6 \times 10^{-5}$ seconds to mitigate bit-flip noise introduced by the hardware~\cite{chandarana2025runtimequantumadvantagedigital,simen2025branch,pfionq}. The constant factors are computed based on the CPU performance on which the heuristics are executed (see Methods).

Finally, the best solution $\mathbf{s}_{\text{ws}}$ from BF-DCQO is duplicated to seed the MTS population. MTS is then executed for $G_{\max}$ generations using a tabu list of length $I_{\text{tabu}}$ and the adaptive mutation schedule described above. The corresponding time is $T_{\text{MTS}} = n_{\text{bitflip}}\times 5.740\times10^{-8}$ seconds. Therefore, the total time of the HSQC reads 
\begin{equation}\label{eq:time_shc}
\begin{aligned}
    T &= \underbrace{n_{\text{sweep}} \times n_{\text{runs}} \times 0.6 \times 10^{-5}}_{\text{SA}} \\
      &\quad + \underbrace{n_{\text{shots}} \times 10^{-4} + n_{\text{sweep}}^{\text{loc}} \times 0.6 \times 10^{-5}}_{\text{BF-DCQO}} \\
      &\quad + \underbrace{n_{\text{bitflip}} \times 5.740 \times 10^{-8}}_{\text{MTS}}~\text{s}.
\end{aligned}
\end{equation}
This hybrid, sequential strategy allows each solver to build upon the progress of the previous stage, enabling deeper exploration of the solution space and faster convergence to high-quality solutions.
\begin{table*}[t]
\caption{\textbf{Per-seed minimum energies and runtimes for MTS, SA, and HSQC.} 
Results correspond to the best-case instance, with each method repeated ten times using different random seeds. 
For MTS, all trials for a given seed were initialized identically and executed with $G_{\rm{max}} = 20000$, employing early stopping upon reaching the ground state. 
For SA, each seed was run using $n^{\rm{SA}}_{\rm{sweep}} = 50000$ sweeps and $n_{\rm{runs}} = 1000$ independent runs. 
For HSQC, the bitstring obtained from BF-DCQO was fixed, and the MTS with initial population from BF-DCQO, was executed with $G_{\rm{max}} = 20000$ and early stopping at the ground state energy. 
The total runtime of HSQC is decomposed into contributions from SA ($T_{\rm{SA}}$), BF-DCQO ($T_{\rm{BFDCQO}}$), and MTS ($T_{\rm{MTS}}$). 
For reference, the exact CPLEX runtime is $T_{\rm{CPLEX}} = 22.896$~s and is independent of seed.}
\label{tab:per_seed_min_compare_extended}
\begin{ruledtabular}
\begin{tabular}{lc cc cc ccccc}
\multirow{2}{*}{Seed} & \multirow{2}{*}{$T_{\rm CPLEX}$ [s]}
& \multicolumn{2}{c}{MTS}
& \multicolumn{2}{c}{SA}
& \multicolumn{5}{c}{HSQC} \\
\cmidrule{3-4} \cmidrule{5-6} \cmidrule{7-11}
&
& Min energy & {$T_{\rm MTS}$ [s]}
& Min energy & {$T_{\rm SA}$ [s]}
& Min energy & {$T_{\rm SA}$ [s]} & {$T_{\rm BFDCQO}$ [s]} & {$T_{\rm MTS}$ [s]} & {$T$ [s]} \\
\midrule
0 & 22.896 & -223.186 &  5.168 & -223.186 & 300.0 & -223.186 & 0.600 & 0.505 & 0.475 & 1.580 \\
1 & 22.896 & -223.186 &  2.274 & -222.900 & 300.0 & -223.186 & 0.600 & 0.505 & 0.177 & 1.282 \\
2 & 22.896 & -219.769 & 18.071 & -222.841 & 300.0 & -223.186 & 0.600 & 0.505 & 0.276 & 1.382 \\
3 & 22.896 & -219.769 & 18.071 & -223.186 & 300.0 & -223.186 & 0.600 & 0.505 & 0.884 & 1.989 \\
4 & 22.896 & -221.434 & 18.071 & -222.841 & 300.0 & -223.186 & 0.600 & 0.505 & 0.470 & 1.575 \\
5 & 22.896 & -222.900 & 18.071 & -223.186 & 300.0 & -223.186 & 0.600 & 0.505 & 0.390 & 1.495 \\
6 & 22.896 & -222.900 & 18.071 & -223.186 & 300.0 & -223.186 & 0.600 & 0.505 & 0.615 & 1.720 \\
7 & 22.896 & -223.186 &  1.187 & -222.674 & 300.0 & -223.186 & 0.600 & 0.505 & 0.715 & 1.821 \\
8 & 22.896 & -223.186 &  0.274 & -222.874 & 300.0 & -223.186 & 0.600 & 0.505 & 0.119 & 1.224 \\
9 & 22.896 & -221.294 & 18.071 & -223.186 & 300.0 & -223.186 & 0.600 & 0.505 & 0.264 & 1.370 
\end{tabular}
\end{ruledtabular}
\end{table*}
For each instance, we execute the first two stages of the HSQC workflow, SA followed by BF-DCQO,  within a fixed time budget, yielding an initial population for MTS. We then run this MTS multiple times and compare its performance against standalone MTS (with a randomly sampled initialization), standard SA, and the exact solver CPLEX. This setup allows us to isolate and assess the contribution of quantum acceleration within the HSQC framework. Regarding MTS, for each run, we fix $I_{\rm{tabu}}=10$, $\mu_{\rm{start}} =0.1$ and $\mu_{\rm{end}}=0.001$ and generation $G_{\rm{max}} \in \{25,50,75,100\}$ as hyperparameters. 

Figure~\ref{fig:AR-MTS} presents a direct comparison between the performance of the MTS algorithm and the HSQC workflow, evaluated using the minimum optimality gap $\bar{\mathcal{G}} = 100\,(\bar{E}_{\textrm{best}} - E_{\textrm{GS}})/|E_{\textrm{GS}}|$, where $E_{\textrm{GS}}$ denotes the ground-state energy obtained from CPLEX~\cite{cplex}. Details of the experimental setup, including statistics for $E_{\rm{min}}$ and the average best energy $\overline{E}_{\text{best}}$, are provided in Table~\ref{tab:mts_bfdcqo} and discussed further in the Methods section.

We observe that all data points fall below the equivalence line $\bar{\mathcal{G}}_{\text{MTS}} = \bar{\mathcal{G}}_{\text{HSQC}}$, clearly demonstrating the consistent advantage of incorporating HSQC. Particularly for smaller generation limits (e.g., $G_{\text{max}} = 25$), data cluster toward the lower right corner, reflecting substantial gains in solution quality due to the HSQC. As $G_{\text{max}}$ increases, the performance gap between MTS and HSQC narrows, with data points approaching the equivalence line. This is expected: more generations allow MTS to further refine solutions, while the relative improvement offered by HSQC diminishes due to saturation and the growing density of local minima near the optimum.

Nonetheless, HSQC maintains superior performance across all tested generation limits, yielding lower $\overline{E}_{\text{best}}$ and, in many instances, smaller $E_{\rm{min}}$. Notably, HSQC successfully recovers the exact ground state for two problem instances with just $G_{\text{max}}=75$, a feat unattained by standalone MTS in any configuration. These findings underscore HSQC’s robustness and efficiency, showing it not only accelerates convergence but also enhances the likelihood of identifying exact or near-exact solutions.

To make a fair runtime-based comparison, we select the best-case HUBO instance and repeat the HSQC (fixed SA + BF-DCQO), MTS, and SA procedures over 10 independent seeds. The corresponding results are summarized in Table~\ref{tab:per_seed_min_compare_extended}. Since HSQC successfully finds the GS for this instance, we use it as a benchmark to assess how long it takes for standalone SA and MTS to achieve the same result.

To this end, we modify the MTS algorithm to allow for a maximum of $G_{\rm{max}} = 20000$ generations, and introduce a stopping criterion based on the target energy $E_{\rm{target}} = -223.1855$, corresponding to the ground-state energy. The MTS column in Table~\ref{tab:per_seed_min_compare_extended} reports the minimum energy reached for each seed, along with the total runtime $T_{\rm{MTS}}$, computed using Eq.~\eqref{eq:time_shc}. If $E_{\rm{target}}$ is reached before exhausting all generations, the algorithm stops early.

For the HSQC protocol, we set the time $T_{\rm{SA}} = 0.6$ seconds, based on $n_{\rm{sweep}}^{\rm{SA}} = 1000$ and $n_{\rm{runs}}^{\rm{SA}} = 100$. The lowest-energy bitstring from SA is then used to initialize the bias fields in BF-DCQO, which is executed using $n_{\rm{shots}} = 5000$ shots. For the local-search, we set $n^{\rm{loc}}_{\rm{sweep}}=900$ sweeps to mitigate bitflip errors. The best result from BF-DCQO is subsequently passed to MTS for initializing the population.

We observe that standalone MTS successfully reaches the ground state in only 4 out of 10 trials, with runtimes reaching up to $T_{\rm{MTS}} \approx 18$ seconds. In contrast, HSQC consistently finds the ground state in all 10 trials, with a significantly lower and consistent total runtime of approximately $T \approx 1.5$ seconds. This clearly demonstrates a runtime advantage of HSQC over standalone MTS.

To complete the comparison, we also evaluate standalone SA. Specifically, we perform 10 independent trials using $n_{\rm{sweep}} = 50000$ and $n_{\rm{runs}} = 1000$, resulting in a total runtime of approximately $T_{\rm{SA}} = 300$ seconds per trial. In 5 out of the 10 trials the ground state is recovered. While this represents a higher success rate than standalone MTS, it comes at a significantly greater computational cost. In contrast, HSQC achieves ground-state solutions in all 10 trials, with a much lower and consistent runtime of approximately 1.5 seconds, demonstrating its runtime advantage over both SA and MTS.

Additionally, we note that the runtime for CPLEX on this instance is approximately 23 seconds. While this includes the time required to prove optimality, it is still longer than the total time required by HSQC to find the ground state. It is important to highlight that HSQC, as a hybrid workflow composed of multiple heuristic components, is inherently probabilistic in nature. Therefore, any comparison with deterministic exact solvers like CPLEX must account for this. However, since HSQC successfully recovers the ground state in all 10 trials for this instance, the runtime comparison remains meaningful and highlights HSQC’s practical efficiency even against exact methods.

\begin{figure}[t]
    \centering 
    \includegraphics[width=\linewidth]{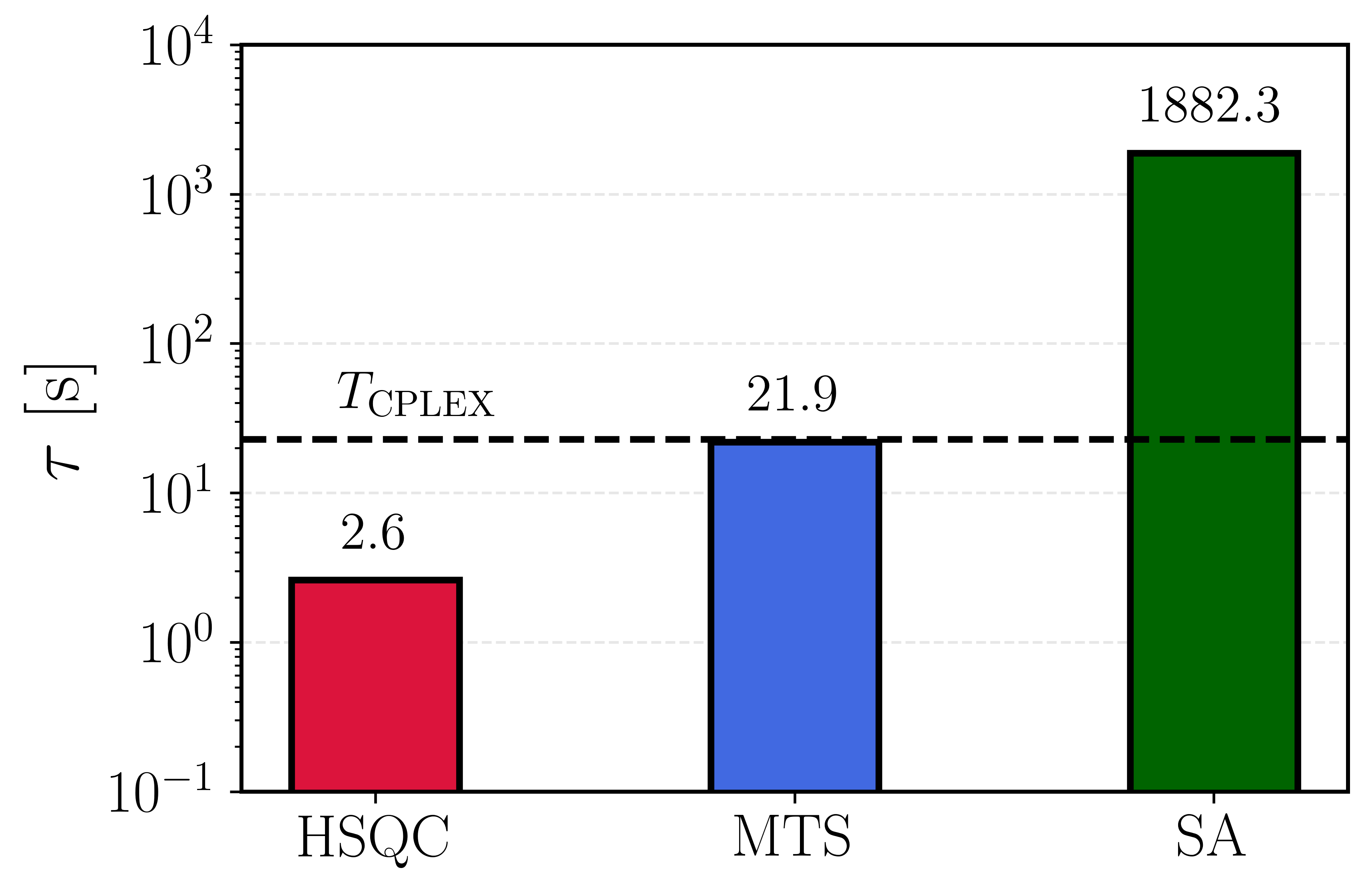}
    \caption{\textbf{Minimum $\tau$ values for various standalone optimization routines for Instance 8.} Comparing minimum $\tau$ values for MTS, HSQC and SA obtained from Table~\ref{tab:mts_shc_tau} and Table.~\ref{tab:sa_tau} with $T_{\rm{CPLEX}} = \SI{22.896}{s}$ value as a reference. }
    \label{fig:Min_tau}
\end{figure}%

As a final benchmark, we evaluate the estimated time for convergence $\tau$ for Instance 8 to compare the performance of HSQC, SA, and MTS heuristics, accounting for their inherently probabilistic nature. We define the convergence time $\tau$ as
\begin{equation}\label{eq:est_time_convergence}
\tau = t_f \frac{\log(1 - 0.99)}{\log(1 - p_{\rm{gs}})},
\end{equation}
where $t_f$ denotes the time per trial, which depends on the specific solver, and $p_{\rm{gs}}$ is the empirical probability of reaching the ground state across all trials. The factor $0.99$ corresponds to the desired confidence level.

For Instance 8, we observe the first occurrences of ground-state solutions starting from $G_{\rm{max}} = 75$ (see Table~\ref{tab:mts_bfdcqo}). Therefore, we perform $100$ independent trials at $G_{\rm{max}} \in \{75, 100, 200, 500\}$ and estimate $p_{\rm{gs}}$ for both MTS and HSQC. Using these probabilities, we compute the corresponding $\tau$ values using Eq.~\eqref{eq:est_time_convergence}. For HSQC, we include the cumulative overhead from the SA and BF-DCQO stages, i.e., $T_{\rm{SA}} + T_{\rm{BFDCQO}} = \SI{1.05}{s}$, in the trial time $t_f$ to ensure a fair runtime comparison.

In Table~\ref{tab:mts_shc_tau}, we report the $p_{\rm{gs}}$, $t_f$ and $\tau$ values for HSQC and MTS. We can observe that the $p_{\rm{gs}}$ values are considerably higher for HSQC than MTS. Therefore, even after adding the constant time, the $\tau$ values of HSQC are much lower than MTS. Specifically for MTS the lowest value we obtained is around $\tau = \SI{21.86}{s}$, whereas for HSQC the minimum value goes down to $\tau=\SI{2.6}{s}$. We run a similar study for the SA case in~\tablabel{tab:sa_tau} where, again, we run 100 independent trials using $n_\text{runs}=1000$ and $n_\text{sweep}\in\{30000, 40000, 50000, 60000, 70000, 100000\}$ and compute the convergence time $\tau$. In this case, we use~\eqlabel{eq:time_shc} to estimate the time per SA trial $t_f$. From the results drawn, the best result obtained comes from the combination of $n_\text{runs}=1000$ and $n_\text{sweep}=50000$, returning $\tau = \SI{1882.30}{s}$, value that is around $700$ times greater than HSQC method. As a side note, we observe that for Instance 8 a saturation point appears when going beyond $n_\text{sweep}=50000$, where increasing the number of sweeps does not help to increase the probability of reaching the ground state energy. The complexity of these instances comes from a distribution that has shown to be challenging for SA~\cite{chandarana2025runtimequantumadvantagedigital}. Fig~\ref{fig:Min_tau}, shows a comparison of the minimum $\tau$ values for HSQC, MTS and SA for Instance 8 with $T_{\rm{CPLEX}}$ for reference as well.

Overall, these results provide compelling evidence that the HSQC paradigm offers a superior balance between solution quality and runtime efficiency. In particular, for dense HUBO problems, HSQC achieves optimal solutions more reliably and faster than standalone classical optimizers such as MTS and SA.

\begin{table}[!tb]
\caption{\textbf{Ground-state probability and runtimes for MTS and HSQC.} Each row corresponds to a maximum generation budget $G_{\rm max}$. Columns report the ground-state success probability $p_{\rm gs}$, the mean runtime $t_f$, and the estimated time $\tau$.}
\label{tab:mts_shc_tau}
\begin{ruledtabular}
\begin{tabular}{c ccc ccc}
\multirow{2}{*}{$G_{\rm{max}}$} & \multicolumn{3}{c}{MTS} & \multicolumn{3}{c}{HSQC} \\
\cmidrule{2-4}\cmidrule{5-7}
& $p_{\rm gs}$ & {$t_f$ [s]} & {$\tau$ [s]} & $p_{\rm gs}$ & {$t_f$ [s]} & {$\tau$ [s]} \\
\midrule
75 & 0.01 & 0.068 & 32.491 & 0.07 & 0.068 & 5.452 \\
100 & 0.02 & 0.091 & \textbf{21.864} & 0.12 & 0.091 & 4.386 \\
200 & 0.02 & 0.181 & 42.447 & 0.41 & 0.181 & 2.688 \\
500 & 0.08 & 0.452 & 26.086 & 0.75 & 0.452 & \textbf{2.608} \\
\end{tabular}
\end{ruledtabular}
\end{table}%
\begin{table}[!tb]
\caption{\textbf{SA ground-state probability and runtimes.} Each row corresponds to a SA sweep budget. Columns report the ground-state success probability $p_{\rm gs}$, the runtime $t_f$, and the estimated time $\tau$.}
\label{tab:sa_tau}
\begin{ruledtabular}
\begin{tabular}{ccccc}
\multirow{2}{*}{$n_{\rm{runs}}$} & \multirow{2}{*}{$n_{\rm{sweep}}$} & \multicolumn{3}{c}{SA} \\
\cmidrule{3-5}
& & $p_{\rm gs}$ & {$t_f$ [s]} & {$\tau$ [s]} \\
\midrule
1000 & 30000 & 0.26 & 180.0 & 2752.96 \\
1000 & 40000 & 0.32 & 240.0 & 2865.82 \\
1000 & 50000 & 0.52 & 300.0 & \textbf{1882.30} \\
1000 & 60000 & 0.35 & 360.0 & 3848.48 \\
1000 & 70000 & 0.37 & 420.0 & 4186.20 \\
1000 & 100000 & 0.36 & 600.0 & 6191.31 \\
\end{tabular}
\end{ruledtabular}
\end{table}

\subsection{SA + BF-DCQO + SA}
\begin{table}[t]
\caption{\textbf{Per-seed minimum energies and runtimes for SA and HSQC for best-performing instance: instance 1.} Hyperparameters are set to $n^{\rm{SA}}_{\rm{sweep}}=10000$ sweeps for SA, $n^{\rm{HSQC}}_{\rm{sweep}}=8167$ for HSQC.}
\label{tab:per_seed_min_compare_extended_sa}
\begin{ruledtabular}
\begin{tabular}{lccc}
Seed & SA & HSQC & $T_\text{SA}=T_\text{HSQC}~$[s] \\ \midrule
0 & -214.6752 & \textbf{-216.1362} & 6.0 \\ 
1 & -215.8490 & \textbf{-215.8609} & 6.0 \\
2 & -215.3549 & \textbf{-215.5230} & 6.0 \\
3 & -214.3795 & \textbf{-216.4535} & 6.0 \\
4 & -215.6728 & \textbf{-215.9188} & 6.0 \\
5 & -214.3795 & \textbf{-215.5230} & 6.0 \\
6 & -215.3549 & \textbf{-217.9757} & 6.0 \\
7 & -215.2783 & \textbf{-215.5230} & 6.0 \\
8 & -214.3083 & \textbf{-215.8609} & 6.0 \\
9 & -214.3515 & \textbf{-215.5230} & 6.0
\end{tabular}
\end{ruledtabular}
\end{table}
Inspired by the previous hybrid combination of methods, we now perform the same study, but now substituting only the MTS part with SA again. Therefore, in this section, we compare the running times and best energies obtained for CPLEX, BF-DCQO, SA, and HSQC, with HSQC results obtained by initializing an SA routine with the 100 lowest energy-valued outcomes of BF-DCQO instead of randomly generated bitstrings. Consequently, the total time of HSQC is now
\begin{equation}\label{eq:time_sa_shc}
\begin{aligned}
    T_\text{HSQC} &= \underbrace{n^\text{SA1}_{\text{sweep}} \times n^\text{SA1}_{\text{runs}} \times 0.6 \times 10^{-5}}_{\text{SA}} \\
      &\quad + \underbrace{n_{\text{shots}} \times 10^{-4} + n_{\text{sweep}}^{\text{loc}} \times 0.6 \times 10^{-5}}_{\text{BF-DCQO}} \\
      &\quad + \underbrace{n^\text{SA2}_{\text{sweep}} \times n^\text{SA2}_{\text{runs}} \times 0.6 \times 10^{-5}}_{\text{SA}}~\text{s}.
\end{aligned}
\end{equation}
We use again as number of sweeps-to-time conversion~\eqlabel{eq:time_shc} and $n_\text{runs}=100$. We use the same parameters as in Sec.~\ref{sec:mts} for the first simulated annealing 

For this variant, we solve the same instances tackled in~\tablabel{tab:mts_bfdcqo} but now substituting the MTS part with SA. In~\tablabel{tab:sa_bfdcqo}, we attach the minimum and mean best energies for SA and HSQC for a different number of sweeps $n^\text{SA2}_\text{sweep}\in\{1000, 2000, 5000, 10000\}$, where the HSQC approach outperforms SA in most of the cases. We also run an intense SA routine with $n_\text{sweep}=10^6$ (leading to $\SI{600}{s}$ of execution), where the ground state energy was found only for three out of the eight instances tested, indicating their complexity for SA. In~\figlabel{fig:sa} we compare the results obtained in~\tablabel{tab:sa_bfdcqo} for all instances in terms of the mean of their best approximation ratios, with most of these values lying in the region where HSQC performed better than SA.

Therefore, rather than exact ground states, for this HSQC variant, we can look instead for better and faster approximate solutions. For that purpose, setting again $n_\text{runs}=100$, we can fix the number of sweeps considered for SA, $n^\text{SA}_\text{sweep}$, and extract the number of sweeps of HSQC, $n^\text{HSQC}_\text{sweep}$, such that $T_\text{SA}=T_\text{HSQC}$. Using the same values of $T_{\{\text{SA},\text{BF-DCQO}\}}$ as in~\tablabel{tab:per_seed_min_compare_extended}, we obtain the relationship $n^\text{HSQC}_\text{sweep} \approx n^\text{SA}_\text{sweep} - 1833$. In~\tablabel{tab:per_seed_min_compare_extended_sa}, we compare for instance 10 the minimum energies obtained when $n^\text{SA}_\text{sweep}=10000$ ($n^\text{HSQC}_\text{sweep}=8167$) is set, which translates into $\SI{6}{s}$ of execution. We can observe that in the ten trials, HSQC returned better solutions in the same amount of time.

\begin{figure}
    \centering
    \includegraphics[width=\linewidth]{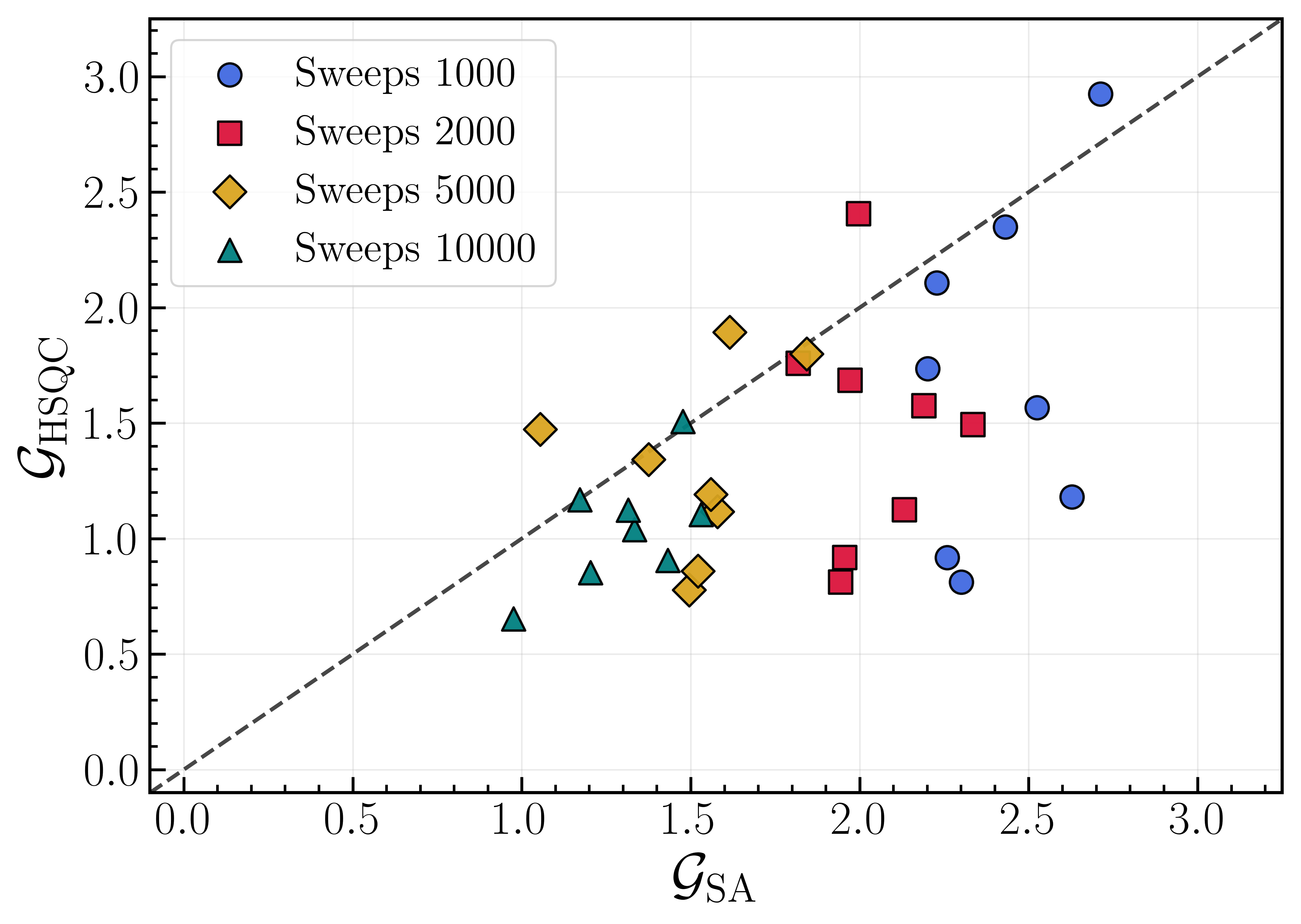}
    \caption{\textbf{Performance comparison between SA and HSQC.} For different number of sweeps, optimality gaps obtained across the eight instances tested in~\tablabel{tab:sa_bfdcqo}. Data points lying in the lower part of the plot mean that the HSQC approach performed better than SA.}\label{fig:sa}
\end{figure}

\section{Discussion}
In this work, we introduce hybrid sequential quantum computing, an optimization framework that integrates simulated annealing, bias-field digitized counterdiabatic quantum optimization, and memetic tabu search, to solve dense higher-order unconstrained binary optimization problems. This approach demonstrates how classical and quantum components can be orchestrated sequentially to exploit their complementary strengths, achieving both rapid convergence and high-quality solutions under the constraints of near-term hardware. Our results reveal that HSQC consistently outperforms standalone SA and MTS in terms of both solution quality and time-to-solution. These findings provide direct evidence of runtime quantum advantage, where incorporating quantum subroutines such as BF-DCQO enables faster and more reliable convergence toward optimal solutions.

A key enabler of this performance is the quantum-generated bitstrings from BF-DCQO to initialize the MTS stage. This steers the classical search toward promising regions of the solution space and away from poor local optima. Across various generation limits, HSQC achieved lower optimality gaps and better minimum energies than standalone MTS, consistently improving both average and worst-case performance. Importantly, HSQC is also well-suited to the limitations and opportunities of upcoming quantum devices. HSQC is designed to find high-quality solutions within fixed time budgets, indicated by the fact that the results can match CPLEX in time-to-solution on certain instances. Its modular design allows each stage to be improved independently, offering flexibility for future enhancements and scalability to larger systems.

Looking ahead, extending HSQC to two-dimensional quantum architectures with a higher degree of connectivity presents a natural and promising next step. Such architectures offer greater connectivity and support for more complex problem instances, increasing the likelihood of runtime separation between classical and quantum solvers. As quantum hardware continues to evolve, we believe HSQC provides a compelling blueprint for industry-relevant hybrid optimization protocols and practical demonstrations of runtime quantum advantage beyond the given benchmarks. To adapt HSQC to industrially relevant problems and specific hardware and classical solvers remains as future work. As quantum hardware scales and more complex hybrid sequences become viable, the disruptive emergence of generative AI and intelligent agents is expected to automate, adapt, and accelerate the HSQC workflows, unlocking its full potential for real-world optimization at unprecedented speed and scale.

\section{Methods}
\subsection{Instance generator}
The instance generator employed in this study builds upon and extends the methodology introduced in Ref.~\cite{chandarana2025runtimequantumadvantagedigital}. Our objective is to generate classically challenging problem instances by co-designing the interaction structure and assigning coupling strengths in a hardware-aware manner.

We begin with empty interaction layers, denoted $\Lambda_2 = \varnothing$ and $\Lambda_3 = \varnothing$, alongside an initial hardware connectivity map $\mathcal{C}^{(0)}$. Using graph-coloring techniques inspired by Ref.~\cite{kim2023evidence}, we identify sets of mutually non-overlapping two- and three-qubit interactions that are compatible with the hardware constraints and can be executed in parallel. These interactions are organized into the families $\mathbb{P}_{2b} = \{\mathbb{P}^{(1)}_{2b}, \ldots, \mathbb{P}^{(R_2)}_{2b}\}$ and $\mathbb{P}_{3b} = \{\mathbb{P}^{(1)}_{3b}, \ldots, \mathbb{P}^{(R_3)}_{3b}\}$, where $R_2$ and $R_3$ denote the number of parallelizable layers for the two- and three-body interactions, respectively.

To construct a specific instance, we select $\rho_{2b}$ and $\rho_{3b}$ layers from the respective families and update the interaction layers as $\Lambda_2 \leftarrow \Lambda_2 \cup \{\mathbb{P}^{(i)}_{2b}\}_{i=1}^{\rho_{2b}}$ and $\Lambda_3 \leftarrow \Lambda_3 \cup \{\mathbb{P}^{(j)}_{3b}\}_{j=1}^{\rho_{3b}}$. Next, we interpret the first two-body layer $\mathbb{P}^{(1)}_{2b}$ as a \textsc{swap} operation, which permutes qubit labels and updates the connectivity map to a new configuration $\mathcal{C}^{(1)}$. This process is repeated iteratively for $n_{\text{swap}}$ rounds, using the leading layer from $\mathbb{P}_{2b}$ in each round to recursively update the connectivity map, as outlined in Algorithm~\ref{alg:swap_layers}.
\begin{figure}[!tb]
\vspace{-2.7mm}
\begin{minipage}[t]{\columnwidth}%
\begin{algorithm}[H]%
\caption{Instance layout generation with swap-based connectivity updates}
\label{alg:swap_layers}
\begin{algorithmic}[1]
\Require Number of swap rounds $n_{\text{swap}}$; initial connectivity map $\mathcal{C}^{(0)}$; integers $\rho_{2b}, \rho_{3b}$; function \textsc{GraphColoring}$(\mathcal{C})$ that returns the families $\mathbb{P}_{2b}=\{\mathbb{P}^{(1)}_{2b},\ldots,\mathbb{P}^{(R_2)}_{2b}\}$ and $\mathbb{P}_{3b}=\{\mathbb{P}^{(1)}_{3b},\ldots,\mathbb{P}^{(R_3)}_{3b}\}$ of mutually non-overlapping two- and three-body interaction layers compatible with $\mathcal{C}$; function \textsc{SwapRegister}$(\mathcal{C}, I)$ that permutes qubit labels in $\mathcal{C}$ according to a set of disjoint pairs $I$ and returns the updated connectivity map.
\State Initialize $\Lambda_2 \gets \varnothing$ and $\Lambda_3 \gets \varnothing$
\State Set $\mathcal{C} \gets \mathcal{C}^{(0)}$
\For{$r = 1$ to $n_{\text{swap}}$}
    \State $(\mathbb{P}_{2b}, \mathbb{P}_{3b}) \gets \textsc{GraphColoring}(\mathcal{C})$ \Comment{$\mathbb{P}_{2b}=\{\mathbb{P}^{(i)}_{2b}\}_{i=1}^{R_2}$, $\mathbb{P}_{3b}=\{\mathbb{P}^{(j)}_{3b}\}_{j=1}^{R_3}$}
    \State $\Lambda_2 \gets \Lambda_2 \cup \{\mathbb{P}^{(i)}_{2b}\}_{i=1}^{\rho_{2b}}$ \Comment{Select $\rho_{2b}$ two-body layers}
    \State $\Lambda_3 \gets \Lambda_3 \cup \{\mathbb{P}^{(j)}_{3b}\}_{j=1}^{\rho_{3b}}$ \Comment{Select $\rho_{3b}$ three-body layers}
    \If{$r < n_{\text{swap}}$}
        \State $\mathcal{C} \gets \textsc{SwapRegister}\!\left(\mathcal{C},\, \mathbb{P}^{(1)}_{2b}\right)$ \Comment{Interpret leading two-body layer as \textsc{swap}}
    \EndIf
\EndFor
\State \Return $(\Lambda_2, \Lambda_3)$
\end{algorithmic}
\end{algorithm}
\end{minipage}
\end{figure}
All simulations in this study use $\mathcal{C}^{(0)}$ corresponding to IBM’s Heron processor, which features a 156-qubit heavy-hexagonal lattice architecture~\cite{chamberland2020topological}.

\subsection{BF-DCQO and circuit decomposition}

The counterdiabatic Hamiltonian is constructed using a first-order nested commutator approximation of the adiabatic gauge potential, resulting in:
\begin{equation}\label{eq:o1}
\begin{aligned}
& H_{cd} = -2\beta_1(t) \Bigg[
\sum_{i=1}^N h_i^{x} h_i^{z}\sigma_i^{y} \\
&\text{ }\text{ } + \sum_{(m,n)\in \mathbb{P}_{2b}} J_{mn} 
   \big( h_m^{x}\sigma_m^{y}\sigma_n^{z} 
       + h_n^{x}\sigma_m^{z}\sigma_n^{y} \big) \\
&\text{ }\text{ } + \sum_{(p,q,r)\in \mathbb{P}_{3b}} \! K_{pqr} \big(
      h_p^{x}\sigma_p^{y}\sigma_q^{z}\sigma_r^{z} 
    + h_q^{x}\sigma_p^{z}\sigma_q^{y}\sigma_r^{z} 
    + h_r^{x}\sigma_p^{z}\sigma_q^{z}\sigma_r^{y} \big)\Bigg],
\end{aligned}
\end{equation}
where the time-dependent scaling function $\beta_1(t)$ is analytically derived in Ref.~\cite{romero2024bias}. The algorithm begins by preparing the ground state of $H_m$ using a product state $\ket{\psi_i} = \bigotimes_{i=1}^N R_y(\theta_i) \ket{0}$, with the angle $\theta_i$ determined by
\[
\theta_i = \tan^{-1} \left( \frac{h^x_i}{h^b_i + \sqrt{(h^b_i)^2 + (h^x_i)^2}} \right).
\]
This state is evolved under $H_{cd}$ in the impulse regime, where time evolution is digitized and includes contributions from one-body, two-body, and three-body interactions.

To construct the quantum circuit, all single-qubit gates are applied in parallel, followed by parallel layers of three-body and then two-body interaction terms. A \textsc{swap} layer is applied at the end of each sequence, and this entire sequence is repeated for $n$ \textsc{swap} layers. At each layer, only $\mathbb{P}_{3b}$ three-body and $\mathbb{P}_{2b}$ two-body terms are applied, following the ordering strategy outlined in Algorithm~\ref{alg:swap_layers}. The deliberate ordering, placing three-body interactions before two-body terms, ensures more efficient gate compilation and helps cancel redundant operations introduced by subsequent swaps.

All the experiments were executed on the IBM Quantum backends \texttt{ibm\_kingston}, \texttt{ibm\_marrakesh}, or \texttt{ibm\_aachen}, depending on queue time and hardware availability~\cite{ibm}. We transpiled the BF-DCQO circuits using Qiskit’s transpiler at optimization level 3~\cite{qiskit}, targeting IBM’s hardware-native gate set $\{\textsc{CZ}, R_z(\theta), \sqrt{X}, X\}$, where $\textsc{CZ} = \mathrm{diag}(1,1,1,-1)$ and $R_z(\theta) = \exp(-i\theta\sigma^z/2)$. To further reduce circuit depth and improve hardware efficiency, we incorporated fractional gates~\cite{ibm_fractional_gates}, natively supported on IBM’s Heron QPUs. Specifically, we used $R_{zz}(\theta) = \exp(-i\theta \sigma^z_0 \sigma^z_1/2)$ for $0 < \theta \leq \pi/2$, and $R_x(\theta) = \exp(-i\theta \sigma^x/2)$ for arbitrary $\theta$, which allowed us to implement entangling operations more compactly.

After each evolution step, $n_{\text{shots}}$ measurements are performed in the computational basis. From the outcome distribution, the $n_{\text{CVaR}}$ lowest-energy bitstrings are selected according to the conditional-value-at-risk (CVaR) approach~\cite{Barkoutsos2020improving,barron2023provableboundsnoisefreeexpectation,romero2024bias}. These configurations inform the update of the bias fields $h^b_i$ for the next iteration of the algorithm. The overall process is repeated for $n_{\text{iter}}$ iterations to progressively refine the solution quality.
\begin{figure}
    \centering
    \includegraphics[width=\linewidth]{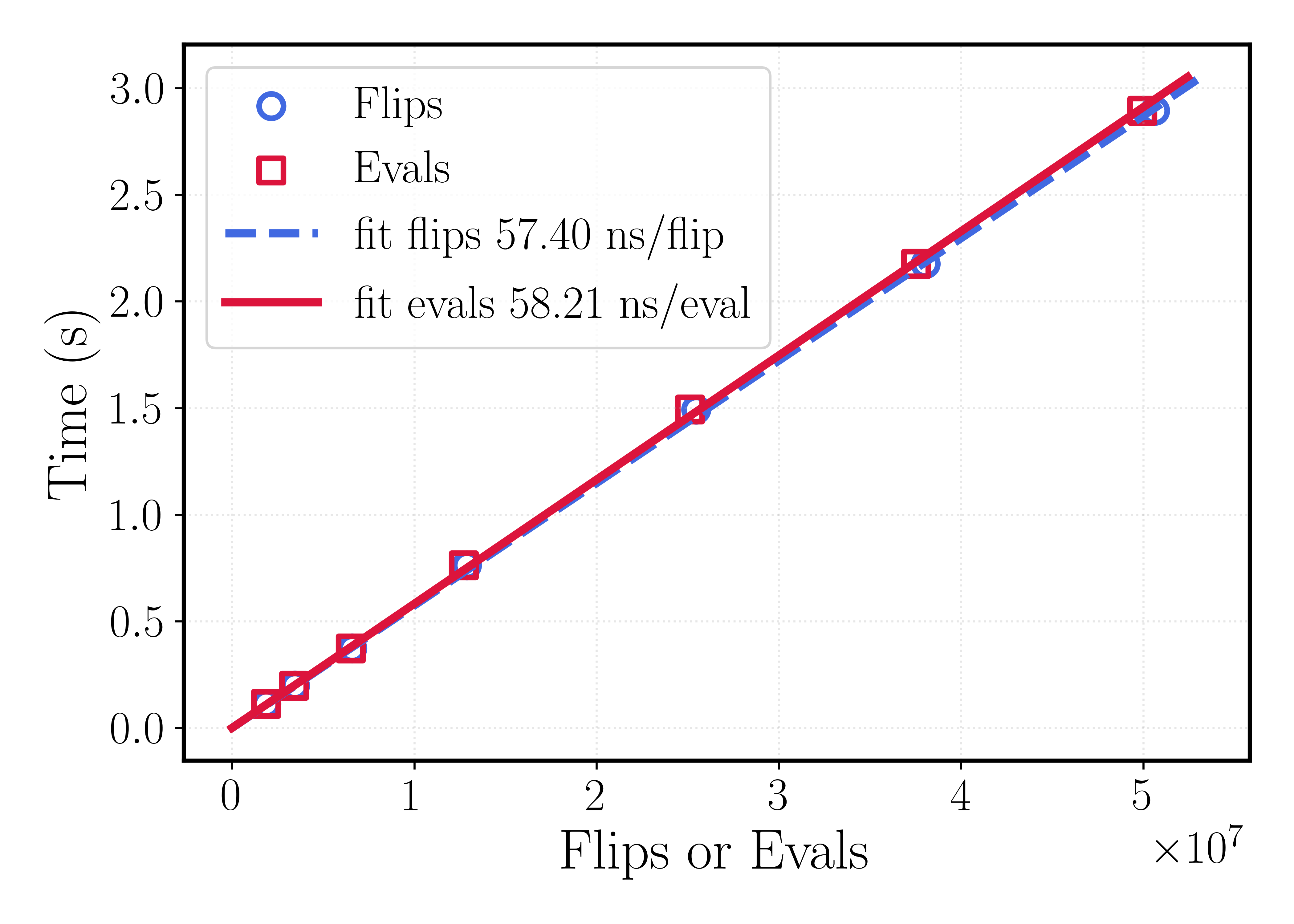}
    \caption{\textbf{Estimating per-flip runtime for MTS.} Total runtime of the Memetic tabu Search (MTS) algorithm as a function of the number of bit-flips (function evaluations). A linear fit is applied to extract the average time per flip, which is subsequently used to estimate MTS runtimes throughout the manuscript.}
    \label{fig:timeestimationmts}
\end{figure}

\subsection{Memetic tabu search}
The goal is to discover a binary string $\mathbf{s} \in \{0,1\}^{n}$ that minimizes an energy functional $E(\mathbf{s})$ defined by an Ising-style problem instance. The \emph{memetic tabu search} (MTS) couples two complementary search strategies: (i) a tabu-guided bit-flip local search that intensively exploits the energy landscape, and (ii) a population-based evolutionary layer that enables global exploration and information exchange among candidate solutions.

The local search component begins from an incumbent string and repeatedly inspects its Hamming-1 neighborhood. At each iteration, every one-bit variant is evaluated, but any configuration found in a short, fixed-length \emph{tabu list} is skipped. The admissible neighbor with the lowest energy becomes the new incumbent, while the previous incumbent is appended to the tabu list. This process continues for a bounded number of iterations or until a user-defined energy target is reached, whichever comes first.

To initialize the population, a set of $P$ candidate strings is generated.  Optionally, a bitstring may be supplied, otherwise, all strings are sampled uniformly at random. Each initial string is immediately refined via the local tabu search, ensuring that the algorithm begins from locally optimized states rather than raw random samples.

%---------------------------------------------
% Algorithm 1 – Local tabu Search
%---------------------------------------------
\begin{figure}[!t]%
\vspace{-2.7mm}
\begin{minipage}[t]{\columnwidth}%
\begin{algorithm}[H]%
\caption{Local tabu Search}
\label{alg:tabu}
\begin{algorithmic}[1]
\Require incumbent $\mathbf{s}$, energy $E(\mathbf{s})$, 
         tabu length $L$,  
         max iterations $I_{\text{tabu}}$,  
         (optional) target $E_{\text{target}}$
\Ensure  locally–optimized string $\mathbf{s}^{\star}$ and energy $E^{\star}$
\State initialize FIFO tabu list $\mathcal{T}\gets\emptyset$
\State $\mathbf{s}^{\star}\gets\mathbf{s}$,\,$E^{\star}\gets E(\mathbf{s})$
\If{$E^{\star}\le E_{\text{target}}$}
    \State \Return $(\mathbf{s}^{\star},E^{\star})$
\EndIf
\For{$k=1$ \textbf{to} $I_{\text{tabu}}$}
    \State $\mathcal{N}\gets$ Hamming-1 neighbors of $\mathbf{s}$ not in $\mathcal{T}$
    \If{$\mathcal{N}=\emptyset$}
        \State \textbf{break}
    \EndIf
    \State choose $\mathbf{v}\in\mathcal{N}$ with minimum $E(\mathbf{v})$
    \State append $\mathbf{s}$ to $\mathcal{T}$; discard oldest if $|\mathcal{T}|>L$
    \State $\mathbf{s}\gets\mathbf{v}$,\,$E(\mathbf{s})\gets E(\mathbf{v})$
    \If{$E(\mathbf{s})<E^{\star}$}
        \State $\mathbf{s}^{\star}\gets\mathbf{s}$,\,$E^{\star}\gets E(\mathbf{s})$
        \If{$E^{\star}\le E_{\text{target}}$}
            \State \Return $(\mathbf{s}^{\star},E^{\star})$
        \EndIf
    \EndIf
\EndFor
\State \Return $(\mathbf{s}^{\star},E^{\star})$
\end{algorithmic}
\end{algorithm}
\end{minipage}%
\end{figure}%
%---------------------------------------------
% Algorithm 2 – Memetic tabu Search
%---------------------------------------------
\begin{figure}[!t]%
\vspace{-2.7mm} % <- tune!!!
\begin{minipage}[t]{\columnwidth}
\begin{algorithm}[H]
\caption{Memetic tabu Search}
\label{alg:mts}
\begin{algorithmic}[1]
\Require bit length $n$, population size $P$, generations $G_{\max}$,  
         tabu iterations $I_{\text{tabu}}$,  
         mutation range $(\mu_{\text{start}},\mu_{\text{end}})$,  
         (optional) bitstring $\mathbf{s}_{\text{ws}}$,  
         (optional) target $E_{\text{target}}$
\Ensure  best string $\mathbf{s}_{\text{best}}$ and energy $E_{\text{best}}$
\State build initial population $\mathcal{P}$:
    \If{bitstring not supplied} 
        \State sample one string uniformly at random
    \EndIf
    \For{each candidate until $|\mathcal{P}|=P$}
        \State refine with \textbf{Local tabu Search}
        \If{target reached} 
            \State \Return best candidate
        \EndIf
    \EndFor
\State $(\mathbf{s}_{\text{best}},E_{\text{best}})\gets$ best member of $\mathcal{P}$
\For{$g=0$ \textbf{to} $G_{\max}-1$}
    \State $\displaystyle \mu_g \gets \mu_{\text{end}} 
            + (\mu_{\text{start}}-\mu_{\text{end}})
              \frac{\ln(G_{\max}+1-g)}{\ln(G_{\max}+1)}$ \Comment{decaying mutation}
    \State $\mathcal{O}\gets\emptyset$ \Comment{offspring set}
    \For{$i=1$ \textbf{to} $P$}
        \State pick two distinct parents uniformly from $\mathcal{P}$
        \State single-point crossover $\rightarrow$ child $\mathbf{c}$
        \State mutate each bit of $\mathbf{c}$ with probability $\mu_g$
        \State refine $\mathbf{c}$ with \textbf{Local tabu Search}
        \If{target reached} 
            \State \Return $(\mathbf{c},E(\mathbf{c}))$
        \EndIf
        \State add $\mathbf{c}$ to $\mathcal{O}$
    \EndFor
    \State $\mathcal{P}\gets$ $P$ best strings from $\mathcal{P}\cup\mathcal{O}$ \Comment{elitist replacement}
    \If{best energy in $\mathcal{P}$ improves $E_{\text{best}}$}
        \State update $(\mathbf{s}_{\text{best}},E_{\text{best}})$
    \EndIf
\EndFor
\State \Return $(\mathbf{s}_{\text{best}},E_{\text{best}})$
\end{algorithmic}
\end{algorithm}
\end{minipage}
\end{figure}

The evolutionary layer proceeds for a prescribed number of generations. In each generation, two parents are selected at random without replacement, combined via single-point crossover, and then subjected to bit-wise mutation. The mutation probability follows a logarithmic decay schedule as shown in \eqlabel{eq:decaying_mutation}. Here, $g$ denotes the current generation and $G_{\max}$ is the total number of generations. This schedule starts aggressively to promote exploration and gradually cools as the search progresses. Each offspring is immediately refined by the local tabu search, yielding a memetically improved individual. Once all offspring have been generated and refined, the union of parents and offspring is ranked by energy, and only the top $P$ solutions are retained. This elitist replacement scheme ensures that the best solution found so far is never lost between generations, while allowing better newcomers to enter the population. The algorithm terminates when any one of the following criteria is met: (i) an individual reaches the desired target energy; (ii) the maximum number of generations is reached; or (iii) no improvement occurs during local refinement for any individual in a generation, indicating stagnation.
\begin{table*}[!tb]
\centering
\caption{\textbf{D-Wave results under different annealing times.} Making use of D-Wave \textsc{Advantage2\_System1.6} quantum annealer, we solve all eight HUBO instances, setting as annealing times $t_a\in\{0.5,20,100,1000,2000\}\,\si{\micro\second}$, using $n_\text{shots}\in\{39920,37030,28570,8000,4440\}$, respetively, to guarantee total sampling times ranging from $5.5-\SI{9.5}{\second}$ on average. For each of them, we attach the minimum energies obtained and optimality gaps $\mathcal{G}$, including the actual ground states (GS) as a reference. Best results per instance are in bold.}\label{tab:dwave_raw}
\begin{ruledtabular}\begin{tabular}{lccccccccccc}
\multirow{2}{*}{Instance} & \multirow{2}{*}{GS} &
\multicolumn{2}{c}{$t_a=\SI{0.5}{\micro\second}$} &
\multicolumn{2}{c}{$t_a=\SI{20}{\micro\second}$} &
\multicolumn{2}{c}{$t_a=\SI{100}{\micro\second}$} &
\multicolumn{2}{c}{$t_a=\SI{1000}{\micro\second}$} &
\multicolumn{2}{c}{$t_a=\SI{2000}{\micro\second}$} \\
\cmidrule{3-4}\cmidrule{5-6}\cmidrule{7-8}\cmidrule{9-10}\cmidrule{11-12}
 &  & $E_{\min}$ & $\mathcal{G}$ (\%) & $E_{\min}$ & $\mathcal{G}$ (\%) & $E_{\min}$ & $\mathcal{G}$ (\%) & $E_{\min}$ & $\mathcal{G}$ (\%) & $E_{\min}$ & $\mathcal{G}$ (\%) \\
\midrule
1 & -218.098 & -161.185 & 26.095 & -166.253 & 23.771 & -170.542 & 21.805 & \textbf{-174.652} & 19.920 & -172.054 & 21.112 \\
2 & -194.771 & -131.297 & 32.589 & -141.377 & 27.414 & -146.813 & 24.623 & \textbf{-147.922} & 24.053 & -147.918 & 24.055 \\
3 & -191.360 & -133.978 & 29.986 & -140.384 & 26.639 & -149.367 & 21.945 & \textbf{-154.865} & 19.071 & -147.603 & 22.866 \\
4 & -242.283 & -172.027 & 28.997 & -182.116 & 24.833 & -188.185 & 22.328 & -185.723 & 23.345 & \textbf{-195.150} & 19.454 \\
5 & -221.722 & -141.397 & 36.228 & \textbf{-163.465} & 26.275 & -160.719 & 27.513 & -158.650 & 28.446 & -157.407 & 29.007 \\
6 & -229.319 & -134.108 & 41.519 & -148.808 & 35.109 & -154.888 & 32.457 & \textbf{-163.039} & 28.903 & -152.001 & 33.716 \\
7 & -219.571 & -131.745 & 39.999 & -145.718 & 33.635 & -153.470 & 30.105 & \textbf{-155.455} & 29.201 & -147.872 & 32.654 \\
8 & -223.186 & -142.857 & 35.992 & -151.062 & 32.316 & -156.517 & 29.871 & \textbf{-158.339} & 29.055 & -154.380 & 30.829 \\
\end{tabular}\end{ruledtabular}
\end{table*}%

To estimate the runtime of the algorithm, we executed MTS with increasing generation limits 
$G_{\rm{max}} \in \{5, 10, 20, 40, 80, 120, 160\}$ on a HUBO instance of size $N = 156$.  
For each run we measured the elapsed time as a function of both the number of bit flips and the number of function evaluations performed.  The results are shown in Fig.~\ref{fig:timeestimationmts}.  By fitting a linear model to these data and extrapolating, we obtained estimates for the time per bit flip and the time per function evaluation.  
Since both quantities grow at nearly identical rates, we define the runtime of MTS as
$T_{\rm{MTS}} \;=\; n_{\rm{bitflip}} \times 5.7 \times 10^{-8}\;\mathrm{s}.$

\subsection{Simulated annealing}
We apply SA directly to the HUBO formulation of~\eqlabel{eq:hubo_ham}. All spins are initialized at random, and an upper bound on the largest single‐spin energy change, $\Delta E^{\max} = \max_i \Delta E_i^{\max}$, is computed. This bound defines the initial temperature, $T_{\text{init}} = \Delta E^{\max}$, while the final temperature is set to $T_{\text{final}} = 0.01\,T_{\text{init}}$. A geometric cooling schedule spans these limits, assigning one temperature per sweep.

Each run comprises $n_{\text{sweep}}$ sweeps. At the start of every sweep, the spin indices are randomly permuted. Visiting spins in this order, we compute the exact energy change $\Delta E$ for a proposed flip and accept the move according to the Metropolis–Hastings rule~\cite{metropolis1953equation,hastings1970monte}. After all sweeps have concluded, the lowest energy encountered in the run is recorded. The entire procedure is repeated for $n_{\text{runs}}$ independent runs, and the best overall configuration is retained. To make full use of available hardware, runs are parallelized across all CPU cores (see Table~\ref{tab:specs}) and time per sweep is set to the value found in Ref.~\cite{chandarana2025runtimequantumadvantagedigital}.

\subsection{Extended Results}

In this section, we start by presenting and discussing extended experimental results using D-Wave after solving the eight instances with their quantum annealing devices directly and their hybrid solvers through the \textsc{LeapHybridBQMSampler} class after HUBO-to-QUBO conversion with the \textsc{dimod} library~\cite{ocean}. In~\tablabel{tab:dwave_raw} we present the obtained results on D-Wave's \textsc{Advantage2\_system1.6} for different annealing times $t_a\in[0.5,2000]\,\si{\micro\second}$, indicating the minimum energies obtained and their corresponding optimality gaps. We set as annealing times $t_a\in\{0.5,20,100,1000,2000\}\,\si{\micro\second}$, using as number of samples $n_\text{shots}\in\{39920,37030,28570,8000,4440\}$, respetively, to guarantee total QPU sampling times ranging approximately from $5.5-\SI{9.5}{\second}$ on average.

For the HUBO-to-QUBO conversion, we set as Lagrange multiplier for the penalty terms $w=1+W_{\max}$, with $W_{\max}$ the largest strength in absolute value of the HUBO problem, which are on average roughly $W_{\max}\approx 6$ for our instances. After conversion, our 156-qubit HUBO instances require 580 qubits in its quadratic form, and posterior embedding on hardware requires 1000 to 1400 qubits, which is between $6\times$ to $9\times$ greater than the number of qubits present in the original HUBO form. We can observe that, as expected, for larger annealing times, better solutions are obtained in general. However, the optimality gaps obtained range from $20\%-35\%$, showcasing that the results obtained on D-Wave hardware are not as good as our HSQC approach. The qubit overhead and the inclusion of constraints not naturally present in the original problem might be some of the reasons behind.
\begin{table*}[!tb]
\centering
\caption{\textbf{D-Wave hybrid solver results.} Across all eight HUBO instances, D-Wave hybrid workflow best solutions and their corresponding optimality gaps (computed as $\mathcal{G}=100(E_{\min}-\mathrm{GS})/|\mathrm{GS}|$) using the Greedy energy as $E_{\min}$ under time limits $T=3$ and $\SI{4}{s}$. Additionally, the actual ground states (GS) and QPU access times $T_\text{QPU}$ are attached. Best results per instance are in bold.}\label{tab:dwave_hybrid}
\begin{ruledtabular}\begin{tabular}{lcccccccccccc}
\multirow{2}{*}{Instance} & \multirow{2}{*}{GS} & \multicolumn{5}{c}{Time limit: $T=\SI{3}{s}$} & \multicolumn{5}{c}{Time limit: $T=\SI{4}{s}$} \\ \cmidrule{3-7}\cmidrule{8-12}
 &  & \multicolumn{1}{c}{Raw} & \multicolumn{1}{c}{Greedy} & \multicolumn{1}{c}{$\mathcal{G}$ (\%)} & \multicolumn{1}{c}{$T_\text{QPU}$~[ms]} & \multicolumn{1}{c}{$T$~[ms]} & \multicolumn{1}{c}{Raw} & \multicolumn{1}{c}{Greedy} & \multicolumn{1}{c}{$\mathcal{G}$ (\%)} & \multicolumn{1}{c}{$T_\text{QPU}$~[ms]} & \multicolumn{1}{c}{$T$~[ms]} \\
\midrule
1 & -218.098 & -186.937 & -195.166 & 10.515 & 51.812  & 2997.692 & -191.620 & \textbf{-208.302} & 4.492 & 155.422 & 4000.550 \\
2 & -194.771 & -165.975 & \textbf{-178.039} & 8.591 & 103.616  & 2991.615 & -164.750 & -173.883 & 10.724 & 155.431 & 3998.272 \\
3 & -191.360 & -160.174 & \textbf{-179.938} & 5.969 & 103.612  & 2989.480 & -163.832 & -178.545 & 6.697 & 155.415 & 3990.059 \\
4 & -242.283 & -219.711 & -226.620 & 6.465 & 103.610  & 3002.142 & -218.911 & \textbf{-226.954} & 6.327 & 155.419 & 3987.376 \\
5 & -221.722 & -187.307 & -193.554 & 12.704 & 103.621  & 2996.628 & -192.608 & \textbf{-200.948} & 9.369 & 155.420 & 3996.789 \\
6 & -229.319 & -197.674 & -208.123 & 9.243 & 103.614  & 2997.377 & -203.128 & \textbf{-217.395} & 5.200 & 155.418 & 3988.413 \\
7 & -219.571 & -190.788 & \textbf{-206.462} & 5.970 & 103.617  & 2992.475 & -189.893 & -201.422 & 8.266 & 155.429 & 3986.460 \\
8 & -223.186 & -191.627 & \textbf{-203.340} & 8.892 & 103.621  & 2989.875 & -191.040 & -203.035 & 9.029 & 155.422 & 4001.009 \\
\end{tabular}\end{ruledtabular}
\end{table*}%
\begin{table*}[!htbp]
\centering
\caption{\textbf{Performance and runtime comparison between MTS and HSQC.} Columns list instance index, ground state energy (GS), BF-DCQO energy, and for several $G_{\max}$ values, the minimum and mean-best energies and optimality gaps from MTS and HSQC. Lower energy and lower gap are indicate better performance and $\mathcal{G} = 100\,(E_{\text{min}} - E_{\text{GS}})/|E_{\text{GS}}|$. Best results are in bold.}\label{tab:mts_bfdcqo}

\begin{ruledtabular}\begin{tabular}{lccccccccc}
\multirow{2}{*}{Instance} & \multirow{2}{*}{GS} & \multirow{2}{*}{BF-DCQO} & \multirow{2}{*}{Gen} & \multicolumn{2}{c}{Minimum energy $E_{\min}$} & \multicolumn{2}{c}{Mean best energy $\overline{E}_{\text{best}}$} & \multicolumn{2}{c}{Optimality gap $\mathcal{G}$ (\%)} \\
\cmidrule{5-6} \cmidrule{7-8} \cmidrule{9-10}
& & & & {MTS} & {HSQC} & {MTS} & {HSQC} & {MTS} & {HSQC} \\
\midrule
\multirow{4}{*}{1} & \multirow{4}{*}{-218.0978} & \multirow{4}{*}{-215.5230} & 25  & -204.0179 & \textbf{-216.5204} & -198.9679 & \textbf{-216.1306} & 6.456 & 0.723 \\
 & & & 50  & -214.8493 & \textbf{-216.5494} & -206.6768 & \textbf{-216.2045} & 1.489 & 0.710 \\
 & & & 75  & -216.2912 & \textbf{-216.3255} & -208.1639 & \textbf{-216.1919} & 0.828 & 0.813 \\
 & & & 100 & -214.7896 & \textbf{-216.5204} & -210.0768 & \textbf{-216.1993} & 1.517 & 0.723 \\
\midrule
\multirow{4}{*}{2} & \multirow{4}{*}{-194.7714} & \multirow{4}{*}{-188.6820} & 25  & -191.4881 & \textbf{-193.2080} & -188.4791 & \textbf{-192.3972} & 1.686 & 0.803 \\
 & & & 50  & -193.4040 & \textbf{-194.7714} & -191.8949 & \textbf{-193.2756} & 0.702 & 0.000 \\
 & & & 75  & -193.8005 & \textbf{-194.7714} & -191.6143 & \textbf{-194.1208} & 0.498 & 0.000 \\
 & & & 100 & -193.9923 & \textbf{-194.7714} & -191.8852 & \textbf{-194.1211} & 0.400 & 0.000 \\
\midrule
\multirow{4}{*}{3} & \multirow{4}{*}{-191.3601} & \multirow{4}{*}{-185.3596} & 25  & -185.3193 & \textbf{-188.3562} & -183.1117 & \textbf{-187.0558} & 3.157 & 1.570 \\
 & & & 50  & -187.2410 & \textbf{-188.7335} & -185.4041 & \textbf{-187.8235} & 2.153 & 1.373 \\
 & & & 75  & -187.9333 & \textbf{-189.3062} & -186.7063 & \textbf{-188.3787} & 1.791 & 1.073 \\
 & & & 100 & \textbf{-190.3348} & -189.4325 & -187.4519 & \textbf{-188.4597} & 0.536 & 1.007 \\
\midrule
\multirow{4}{*}{4} & \multirow{4}{*}{-242.2827} & \multirow{4}{*}{-238.0770} & 25  & -239.2863 & \textbf{-239.7158} & -234.3365 & \textbf{-239.0612} & 1.237 & 1.059 \\
 & & & 50  & \textbf{-240.6513} & -240.5283 & -237.3003 & \textbf{-240.0373} & 0.673 & 0.724 \\
 & & & 75  & -241.4677 & \textbf{-241.5935} & -237.7555 & \textbf{-240.5258} & 0.336 & 0.284 \\
 & & & 100 & \textbf{-240.8505} & -240.5830 & -238.8252 & \textbf{-240.3679} & 0.591 & 0.702 \\
\midrule
\multirow{4}{*}{5} & \multirow{4}{*}{-221.7218} & \multirow{4}{*}{-219.9218} & 25  & -211.7941 & \textbf{-219.9219} & -206.2524 & \textbf{-219.9219} & 4.478 & 0.812 \\
 & & & 50  & -213.0573 & \textbf{-220.3947} & -210.2753 & \textbf{-219.9691} & 3.908 & 0.599 \\
 & & & 75  & -215.9287 & \textbf{-219.9219} & -212.8430 & \textbf{-219.9219} & 2.613 & 0.812 \\
 & & & 100 & -216.1129 & \textbf{-221.0083} & -214.4136 & \textbf{-220.0305} & 2.530 & 0.322 \\
\midrule
\multirow{4}{*}{6} & \multirow{4}{*}{-229.3192} & \multirow{4}{*}{-225.7244} & 25  & -221.2532 & \textbf{-229.1247} & -213.4332 & \textbf{-228.3942} & 3.517 & 0.085 \\
 & & & 50  & -227.8142 & \textbf{-229.1247} & -221.4388 & \textbf{-228.7137} & 0.656 & 0.085 \\
 & & & 75  & -228.0460 & \textbf{-229.1247} & -223.0950 & \textbf{-228.8468} & 0.555 & 0.085 \\
 & & & 100 & -228.8585 & \textbf{-229.1626} & -223.7416 & \textbf{-229.0780} & 0.201 & 0.068 \\
\midrule
\multirow{4}{*}{7} & \multirow{4}{*}{-219.5705} & \multirow{4}{*}{-210.6941} & 25  & -212.7764 & \textbf{-215.3337} & -209.1954 & \textbf{-213.8719} & 3.094 & 1.930 \\
 & & & 50  & \textbf{-216.6372} & -215.8933 & -212.7452 & \textbf{-214.7094} & 1.336 & 1.675 \\
 & & & 75  & \textbf{-219.0577} & -216.0782 & -213.4183 & \textbf{-214.9631} & 0.234 & 1.591 \\
 & & & 100 & \textbf{-217.4918} & -216.8694 & -214.4788 & \textbf{-215.5784} & 0.947 & 1.230 \\
\midrule
\multirow{4}{*}{8} & \multirow{4}{*}{-223.1855} & \multirow{4}{*}{-221.1362} & 25  & -218.6106 & \textbf{-222.9956} & -216.3979 & \textbf{-221.6585} & 2.050 & 0.085 \\
 & & & 50  & -220.1601 & \textbf{-222.9956} & -217.5255 & \textbf{-221.9285} & 1.355 & 0.085 \\
 & & & 75  & -221.2916 & \textbf{-223.1856} & -219.2352 & \textbf{-222.0257} & 0.848 & 0.000 \\
 & & & 100 & -220.5968 & \textbf{-223.1856} & -219.1314 & \textbf{-222.3384} & 1.160 & 0.000
\end{tabular}\end{ruledtabular}

\end{table*}%%%%%%%
\begin{table*}[!htbp]
\centering
\caption{\textbf{Performance and runtime comparison between SA and HSQC.} Columns list instance index, ground state energy (GS), SA energy ($\SI{600}{\second}$), BF-DCQO energy, and for each number of sweeps the minimum and mean-best energies and optimality gaps from SA and HSQC. Lower energy and lower gap are indicate better performance and $\mathcal{G} = 100\,(E_{\text{min}} - E_{\text{GS}})/|E_{\text{GS}}|$. Best results are in bold.}\label{tab:sa_bfdcqo}

\begin{ruledtabular}\begin{tabular}{lcccccccccc}
\multirow{2}{*}{Instance} & \multirow{2}{*}{GS} & \multirow{2}{*}{BF-DCQO} & \multirow{2}{*}{SA ($\SI{600}{\second}$)} & \multirow{2}{*}{Sweeps} & \multicolumn{2}{c}{Minimum energy $E_{\min}$} & \multicolumn{2}{c}{Mean best energy $\overline{E}_{\text{best}}$} & \multicolumn{2}{c}{Optimality gap $\mathcal{G}$ (\%)} \\
\cmidrule{6-7} \cmidrule{8-9} \cmidrule{10-11}
& & & & & {SA} & {HSQC} & {SA} & {HSQC} & {SA} & {HSQC} \\
\midrule
\multirow{4}{*}{1} & \multirow{4}{*}{-218.0978} & \multirow{4}{*}{-215.5230} & \multirow{4}{*}{-218.0978} & 1000 & -215.2862 & \textbf{-215.5230} & -212.3657 & \textbf{-215.5230} & 1.289 & 1.181 \\
& & & & 2000 & -215.2182 & \textbf{-216.7411} & -213.4481 & \textbf{-215.6448} & 1.320 & 0.622 \\
& & & & 5000 & -215.9142 & \textbf{-216.0701} & -214.6536 & \textbf{-215.6611} & 1.001 & 0.930 \\
& & & & 10000 & -215.9533 & \textbf{-217.6037} & -214.9753 & \textbf{-216.1209} & 0.983 & 0.227 \\ \midrule
\multirow{4}{*}{2} & \multirow{4}{*}{-194.7714} & \multirow{4}{*}{-188.6821} & \multirow{4}{*}{-194.1150} & 1000 & \textbf{-192.4975} & -191.8835 & -190.4311 & \textbf{-190.6678} & 1.168 & 1.483 \\
& & & & 2000 & -191.8591 & \textbf{-193.3848} & -191.2304 & \textbf{-191.3469} & 1.495 & 0.712 \\
& & & & 5000 & \textbf{-193.7014} & -193.2726 & -192.0922 & \textbf{-192.1573} & 0.549 & 0.770 \\
& & & & 10000 & -192.8074 & \textbf{-194.2513} & -192.1736 & \textbf{-192.7463} & 1.008 & 0.267 \\ \midrule
\multirow{4}{*}{3} & \multirow{4}{*}{-191.3602} & \multirow{4}{*}{-185.3597} & \multirow{4}{*}{-190.9482} & 1000 & -188.3781 & \textbf{-188.8758} & -186.7092 & \textbf{-186.8656} & 1.558 & 1.298 \\
& & & & 2000 & -188.9155 & \textbf{-190.1625} & -187.1693 & \textbf{-188.3467} & 1.278 & 0.626 \\
& & & & 5000 & \textbf{-189.2850} & -188.4505 & \textbf{-188.2682} & -187.7365 & 1.084 & 1.521 \\
& & & & 10000 & \textbf{-190.2138} & -189.7987 & \textbf{-188.5346} & -188.4772 & 0.599 & 0.816 \\ \midrule
\multirow{4}{*}{4} & \multirow{4}{*}{-242.2827} & \multirow{4}{*}{-238.0770} & \multirow{4}{*}{-241.7211} & 1000 & \textbf{-239.3209} & -238.0770 & -236.9490 & \textbf{-238.0770} & 1.223 & 1.736 \\
& & & & 2000 & \textbf{-240.1531} & -238.8498 & -237.5054 & \textbf{-238.2007} & 0.879 & 1.417 \\
& & & & 5000 & \textbf{-241.5584} & -240.6580 & -238.5045 & \textbf{-239.3972} & 0.299 & 0.671 \\
& & & & 10000 & \textbf{-240.6361} & -240.2790 & -239.0974 & \textbf{-239.5613} & 0.680 & 0.827 \\ \midrule
\multirow{4}{*}{5} & \multirow{4}{*}{-221.7219} & \multirow{4}{*}{-219.9219} & \multirow{4}{*}{-221.7219} & 1000 & -219.7765 & \textbf{-219.9219} & -216.6210 & \textbf{-219.9219} & 0.877 & 0.812 \\
& & & & 2000 & -218.6638 & \textbf{-219.9219} & -217.4121 & \textbf{-219.9219} & 1.379 & 0.812 \\
& & & & 5000 & -219.7485 & \textbf{-220.3197} & -218.4066 & \textbf{-219.9977} & 0.890 & 0.632 \\
& & & & 10000 & \textbf{-221.3394} & -220.6576 & -219.5597 & \textbf{-220.2752} & 0.173 & 0.480 \\ \midrule
\multirow{4}{*}{6} & \multirow{4}{*}{-229.3193} & \multirow{4}{*}{-225.7245} & \multirow{4}{*}{-229.3127} & 1000 & \textbf{-226.5969} & -225.7245 & -223.5287 & \textbf{-225.7245} & 1.187 & 1.568 \\
& & & & 2000 & -225.8982 & \textbf{-226.4108} & -223.9656 & \textbf{-225.8952} & 1.492 & 1.268 \\
& & & & 5000 & \textbf{-228.5780} & -227.5673 & \textbf{-226.9006} & -225.9407 & 0.323 & 0.764 \\
& & & & 10000 & \textbf{-228.0932} & -228.0684 & -226.6325 & \textbf{-226.6413} & 0.535 & 0.546 \\ \midrule
\multirow{4}{*}{7} & \multirow{4}{*}{-219.5705} & \multirow{4}{*}{-210.6941} & \multirow{4}{*}{-219.4862} & 1000 & \textbf{-215.9558} & -215.2018 & \textbf{-213.6149} & -213.1492 & 1.646 & 1.990 \\
& & & & 2000 & \textbf{-217.3946} & -216.6785 & \textbf{-215.1870} & -214.2861 & 0.991 & 1.317 \\
& & & & 5000 & \textbf{-217.6086} & -216.3228 & -215.5243 & \textbf{-215.6178} & 0.894 & 1.479 \\
& & & & 10000 & -218.0402 & \textbf{-218.8007} & -216.2090 & \textbf{-217.1442} & 0.697 & 0.351 \\ \midrule
\multirow{4}{*}{8} & \multirow{4}{*}{-223.1856} & \multirow{4}{*}{-221.1363} & \multirow{4}{*}{-223.1856} & 1000 & -220.7617 & \textbf{-221.1363} & -218.1452 & \textbf{-221.1363} & 1.086 & 0.918 \\
& & & & 2000 & -220.4307 & \textbf{-221.1363} & -218.8203 & \textbf{-221.1363} & 1.234 & 0.918 \\
& & & & 5000 & -221.0350 & \textbf{-221.7688} & -219.7895 & \textbf{-221.2686} & 0.964 & 0.635 \\
& & & & 10000 & -221.2851 & \textbf{-222.9956} & -220.5006 & \textbf{-221.2825} & 0.852 & 0.085 
\end{tabular}
\end{ruledtabular}
\end{table*}
 
Given that D-Wave also offers hybrid classical-quantum solutions for a wide range of optimization problems~\cite{dwave2020hybrid}, we again solve the eight instances to compare our HSQC results with their hybrid workflow. Essentially, given a time limit $T$ (in seconds) to solve the input problem, their service starts by running in parallel different heuristic methods on CPUs and GPUs, whose set of solutions might be used to further refine them by means of quantum resources in a later stage~\cite{dwave2021cqm}. Despite the classical algorithms involved and the recipe behind their decomposition pipeline being proprietary, previous documentation and technical reports indicate that \textsc{qbsolv}~\cite{dwave2017partitioning} and \textsc{Kerberos}~\cite{kerberos} have been taken into account, whose heuristics involve tabu search and simulated annealing. 

In~\tablabel{tab:dwave_hybrid} we present the obtained results across all instances using as time limits $T=3$ and $\SI{4}{s}$ employing again the same HUBO-to-QUBO conversion procedure as for the~\tablabel{tab:dwave_raw} results. For this case, we observe in general that the larger the time limit set the better solutions obtained. However, despite the hybrid solutions are lower in energy when compared to the standalone quantum annealing ones (mostly due to the additional classical heuristics), the obtained results are still worse than the ones reported for our HSQC approach.

Apart from this, we present extended experimental results supporting Fig.~\ref{fig:AR-MTS} and Fig.~\ref{fig:sa}. We evaluate eight HUBO instances using both variants of the HSQC pipeline:  SA+BF-DCQO+MTS and SA+BF-DCQO+SA. Each instance is executed ten times, and we report the minimum energy $E_{\rm{min}}$ across all trials, as well as the mean of the best energies obtained in each trial. In addition to the average optimality gap $\bar{\mathcal{G}}$, which is visualized in Fig.~\ref{fig:AR-MTS} and Fig.~\ref{fig:sa}, we also provide the optimality gap $\mathcal{G}$ computed from the minimum energy achieved. Table~\ref{tab:mts_bfdcqo} summarizes results for the MTS variant, showing performance as a function of increasing generation limits $G_{\rm{max}} \in \{25, 50, 75, 100\}$. In contrast, Table~\ref{tab:sa_bfdcqo} presents results for the SA variant, reporting performance for increasing sweep counts $n_{\rm{sweep}} \in \{1000, 2000, 5000, 10000\}$. For context, we also include SA results at $n^{\rm{SA}}_{\rm{sweep}} = 100{,}000$ and $n^{\rm{SA}}_{\rm{runs}} = 1000$, corresponding to a reference runtime of $T_{\rm{SA}} = 600$~seconds.

\begin{acknowledgments}
    We thank Michael Falkenthal and Sebastian Wagner for their help with running the experiments via the PLANQK platform. We acknowledge the use of IBM Quantum services for this work. The views expressed are those of the authors and do not reflect the official policy or position of IBM or the IBM Quantum team.
\end{acknowledgments}

\bibliography{ref.bib}

\end{document}